# On Fractional Kinetic Equations Involving Srivastava Polynomial

[1]**Komal Prasad Sharma, [2]Alok Bhargava\* and Omprakash Saini**
[1]Department of Mathematics and Statistics, NIMS University, Jaipur, (India)
[2]Department of Mathematics and Statistics, Manipal University Jaipur, Jaipur (India)
[3]Department of Mathematics, Sangam University, Bhilwara (India)
[1]keshav4maths@gmail.com, [2]alok.bhargava@jaipur.manipal.edu, [3]opsainisikar@gmail.com

**Abstract**
Kinetic equations hold a very important place in physics and further their fractional generalization enhances the scope of their applicability and significance in describing the continuity of motion in materials. After the development of generalized form of fractional kinetic equations, many researchers proffered several new forms of these equations and found their solutions by different techniques. In this work, we have proposed some novel generalised fractional kinetic equations involving the Srivastava polynomial and, by applying the Laplace transform approach, their solutions are calculated. Further, to study the behaviour of these, numerical and graphical interpretation of the solutions are also provided.

## 1. Introduction

Kinetic equations play crucial role in mathematical physics and natural sciences which describe the continuity of motion of material. Considering its growing number of scientific applications over the past three decades, these become increasingly appealing to scholars in this field. Quickly, kinetic theory found its way into other areas like plasma physics and semiconductors. Kinetic equations have been utilised in several areas of applied mathematics to characterize the behaviour of massive interacting particle systems. The classical applications of kinetic equations include the motion of atoms and molecules in a vacuum, the flow of ions in a plasma, or the charge transfer in a semiconductor crystal. Classical kinetic models have been investigated for quite some time. Such include the Boltzmann equation and the Vlasov-Poisson equation.

Over the past few decades, it has been demonstrated that fractional order differential equations are increasingly important in a number of physical scientific and engineering fields. Due to the significance in areas of applied science like astrophysics, dynamical systems, control systems, and mathematical physics, the solution of the fractional kinetic equations has attracted the interest of many researchers in recent years. Some physical phenomena have been explained using fractional order kinetic equations, the description of continuity of motion of substances is the best example of kinetic equations. In the past few decades, fractional kinetic equations (FKE) have been widely and effectively used to describe and solve several significant physics and astronomy problems. As a result, the related literature contains a significant number of articles about the solutions of these equations.

Considering the significance of kinetic equations and the need to expand their applications in science and mathematics, Haubold and Mathai [5] developed a fractional generalisation of the kinetic equation and Saxena and Kalla [11] explored and described the FKE after conducting more research on the topic as following.

$$N(t) - N_0 f(t) = -c^\alpha \, _0D_t^{-\alpha} \, N(t), \Re(\alpha) > 0 \tag{1.1}$$

where $_0D_t^{-\alpha}$ is the fractional integral operator, defined as

$$_0D_t^{-\alpha} f(x) = \frac{1}{\Gamma(\alpha)} \int (t-u)^\alpha f(u) du, x > 0, \Re(\alpha) > 0 \tag{1.2}$$

where $N(t)$ the numerical density of species at time t is represented by N(t), the number of densities at the time $t = 0$, represent by $N_0$, $f(t) \in L(0, \infty)$, $c$ is the constant.

In addition, the number the researcher like as Agarwal and Bhargava [2], Abbas et al. [1], Ahmed et al. [3], Gupta et al. [4], Oldham and Spanier [8], Özarslan [9], Saichev, A. I. et al. [10], Haubold and Mathai [5], Miller and Ross [6], Saxena and Kalla [11], Saxena et al. [12], Sharma and Bhargava [15-17], Sharma et al. [18], Suthar et al. [20], etc. are investigated and developed several fascinating conclusions of considerable importance regarding the extension of fractional kinetic equations pertaining to special functions.

Further, we provide a generalized version of the fractional kinetic equation connected to the generalized Srivastava polynomial and its fractional derivatives and find a solution using the approach of the Laplace transform technique. The Laplace transform [13,14] of the Riemann–Liouville fractional integral operator [6] over a function $f(t)$ is defined as

$$L\{_0D_t^{-\alpha} f(t); s\} = s^{-\alpha} F(s)$$

The Srivastava polynomials [19] defined as

$$S_w^p(\xi) = \sum_{k=0}^{[w/p]} \frac{(-w)_{pk}}{k!} A_{w,k}(\xi)^k, p \epsilon \mathbb{N}, k \in \mathbb{N}_0$$

(1.3)

Where $\mathbb{N}_0 = \mathbb{N} \cup \{0\}$ and the coefficients $A_{w,k}$ ($w, k \in \mathbb{N}_0$) $\geq 0$ are arbitrary constants real or complex.

**Fractional Derivative of generalized Srivastava polynomials**

$$_0D_t^{\lambda}\{ S_w^p(t)\} = \sum_{k=0}^{[w/p]} \frac{(-w)_{pk}\Gamma(k+1)}{k!\Gamma(k-\lambda+1)} A_{w,k}(t)^{k-\lambda}$$

(1.4)

**Laplace Transform of Fractional Derivative of generalized Srivastava polynomials**

$$L[_0D_t^{\lambda}\{ S_w^p(t)\}] = \sum_{k=0}^{[w/p]} \frac{(-w)_{pk}\Gamma(k+1)}{k!} A_{w,k}(s)^{-(k-\lambda+1)}$$

(1.5)

2. **Main Results**

**Theorem1:** If $v > 0, c > 0, p \epsilon \mathbb{N}, k \in \mathbb{N}_0, A_{w,r} \geq 0, \mathbb{N}_0 = \mathbb{N} \cup \{0\}$, then the solution of the FKE

$$N(t) - N_0\{ S_w^p(t)\} = -c^\alpha{}_0D_t^{-\alpha} N(t)$$

(2.1)

is given by

$$N(t) = N_0 \sum_{k=0}^{[w/p]} \frac{(-w)_{pk}\Gamma(\alpha k + 1)}{k!} A_{w,k}(t)^k E_{\alpha,k+1}(-c^\alpha t^\alpha)$$

(2.2)

where $E_{v,k+1}(.)$ is the Mittag-Leffler function [7].

**Proof:** Using Laplace Transform on (2.1), we have

$$\mathcal{N}(s) = N_0 \sum_{k=0}^{[w/p]} \frac{(-w)_{pk}\Gamma(\alpha k + 1)}{k!} A_{w,k}(s)^{-(k+1)} \times (1 + c^\alpha s^{-\alpha})^{-1}$$

$$= N_0 \sum_{k=0}^{[w/p]} \frac{(-w)_{pk}\Gamma(\alpha k + 1)}{k!} A_{w,k}(s)^{-(k+1)} \sum_{n=0}^{\infty}(-c^\alpha s^{-\alpha})^n$$

$$= N_0 \sum_{k=0}^{[w/p]} \frac{(-w)_{pk}\Gamma(\alpha k + 1)}{k!} A_{w,k} \sum_{n=0}^{\infty}(-c^\alpha)^n (s)^{-(k+1+\alpha n)}$$

(2.3)

Using inverse Laplace Transform of (2.3)

$$N(t) = N_0 \sum_{k=0}^{[w/p]} \frac{(-w)_{pk}\Gamma(\alpha k + 1)}{k!} A_{w,k}(t)^k \sum_{n=0}^{\infty} \frac{(-c^\alpha t^\alpha)^n}{\Gamma(\alpha n + k + 1)}$$

$$= N_0 \sum_{k=0}^{[w/p]} \frac{(-w)_{pk}\Gamma(\alpha k + 1)}{k!} A_{w,k}(t)^k E_{\alpha,k+1}(-c^\alpha t^\alpha)$$

**Numerical and Graphical Interpretation**

By assuming $\alpha$ as constant and experimenting with different values of $t$, we get different values of $N(t)$, for (2.2), which are interpreted in the following Table 2.1 and the graphs 1(a), 1(b) are presented to exhibit the behaviour of the result obtained.

**Table 2.1: The values of $N(t)$ with fix $\alpha$**

| $t$ | Fix $\alpha = 0.1$ $N(t)$ | Fix $\alpha = 0.5$ $N(t)$ | Fix $\alpha = 0.9$ $N(t)$ | Fix $\alpha = 1.3$ $N(t)$ |
|---|---|---|---|---|
| 0 | 0 | 0 | 0 | 0 |
| 0.2 | -1.43513E-07 | -0.238526181 | -0.217842834 | -1.605299442 |
| 0.4 | -0.000298869 | -43.69709741 | -42.2789245 | -191.7275747 |
| 0.6 | -0.026303068 | -928.528993 | -921.8483677 | -3145.979887 |
| 0.8 | -0.633838855 | -8177.662845 | -8213.37053 | -22903.27976 |
| 1 | -7.510926841 | -44481.2612 | -44817.53611 | -106823.3185 |
| 1.2 | -56.80806919 | -178460.2679 | -179356.3817 | -375952.1063 |
| 1.4 | -315.1661277 | -580517.9733 | -579539.5531 | -1089454.94 |
| 1.6 | -1393.604616 | -1619958.489 | -1601291.308 | -2738589.622 |
| 1.8 | -5181.632718 | -4021558.205 | -3926038.225 | -6175656.57 |
| 2 | -16803.48438 | -9104456.68 | -8760155.746 | -12783364.76 |
| 2.2 | -48782.38444 | -19131658.69 | -18112186.93 | -24690087.39 |
| 2.4 | -129246.7848 | -37804680.69 | -35165100.95 | -45036545.3 |
| 2.6 | -317117.2252 | -70945005.97 | -64762569.32 | -78299547.98 |
| 2.8 | -728785.315 | -127412110.6 | -114030019.5 | -130678539.4 |
| 3 | -1582969.24 | -220316967.5 | -193153090.5 | -210550839.5 |
| 3.2 | -3273357.326 | -368600216.2 | -316338075.6 | -329001639.6 |
| 3.4 | -6482510.656 | -599055689.5 | -502981003.9 | -500435004.1 |
| 3.6 | -12355491.42 | -948892818.3 | -779074191.7 | -743272345.6 |
| 3.8 | -22758046.8 | -1468945687 | -1178881404 | -1080745087 |
| 4 | -40652169.65 | -2227652307 | -1746915207 | -1541788483 |

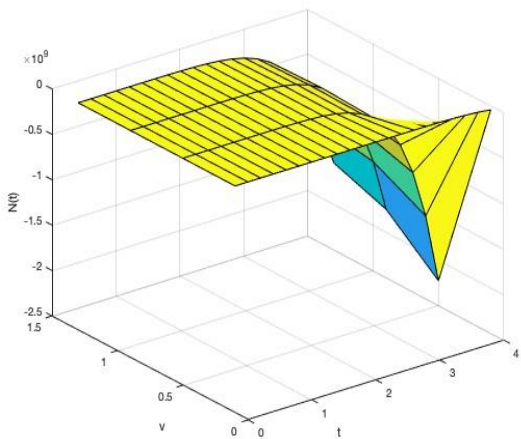 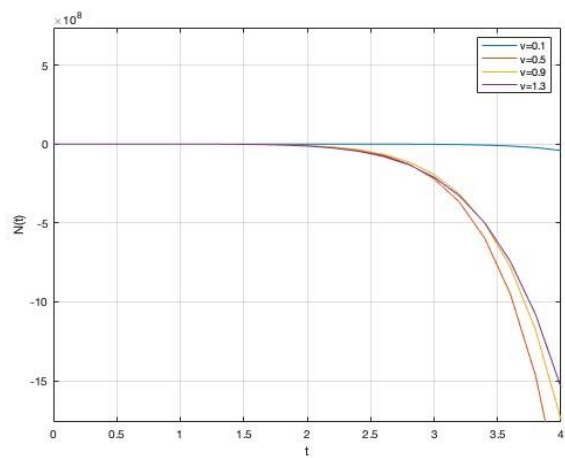

**Figure 2.1(a): 3D graph for (2.2)**             **2.1(b): 2D graph for (2.2)**

**Theorem-2:** If $\alpha > 0, c > 0, p \epsilon \mathbb{N}, k \in \mathbb{N}_0, A_{w,k} \geq 0, \mathbb{N}_0 = \mathbb{N} \cup \{0\}$ then the solution of

$$N(t) - N_0\{S_w^p(c^\alpha t^\alpha)\} = -d^\alpha {}_0D_t^{-\alpha} N(t) \tag{2.4}$$

is given as

$$N(t) = N_0 \sum_{k=0}^{[w/p]} \frac{(-w)_{pk}\Gamma(\alpha k + 1)}{k!} A_{w,k} (ct)^{\alpha k} E_{\alpha,(\alpha k+1)}(-d^\alpha t^\alpha) \tag{2.5}$$

**Proof:** Using Laplace Transform on (2.4), we have

$$\mathcal{N}(s) = N_0 \sum_{k=0}^{[w/p]} \frac{(-w)_{pk}\Gamma(\alpha k + 1)}{k!} A_{w,k} \, s^{-(\alpha k+1)}(c)^{\alpha k}(1 + d^\alpha s^{-\alpha})^{-1}$$

$$= N_0 \sum_{k=0}^{[w/p]} \frac{(-w)_{pk}\Gamma(\alpha k + 1)}{k!} A_{w,k} (c)^{\alpha k} \sum_{n=0}^{\infty} (-d^\alpha s^{-\alpha})^n (s)^{-(\alpha k+1)}$$

$$= N_0 \sum_{k=0}^{[w/p]} \frac{(-w)_{pk}\Gamma(\alpha k + 1)}{k!} A_{w,k} (c)^{\alpha k} \sum_{n=0}^{\infty} (-d^\alpha)^n (s)^{-(\alpha k+1+\alpha n)}$$

Using inverse Laplace Transform, we have

$$N(t) = N_0 \sum_{k=0}^{[w/p]} \frac{(-w)_{pk}\Gamma(\alpha k + 1)}{k!} A_{w,k} (c)^{\alpha k} \sum_{n=0}^{\infty} (-d^\alpha)^n \frac{(t)^{\alpha(k+n)}}{\Gamma(\alpha(n+k)+1)}$$

$$= N_0 \sum_{k=0}^{[w/p]} \frac{(-w)_{pk}\Gamma(\alpha k + 1)}{k!} A_{w,k} (ct)^{\alpha k} E_{\alpha,(\alpha k+1)}(-d^\alpha t^\alpha)$$

**Numerical and Graphical Interpretation**

By assuming $\alpha$ as constant and experimenting with different values of $t$, we get different values of $N(t)$, for (2.5), which are interpreted in the following table 2.2 and the graphs 2(a), 2(b) are presented to exhibit the behaviour of the result obtained.

**Table 2.2: The values of $N(t)$ with fix $\alpha$**

| $t$ | Fix $\alpha = 0.1$ | Fix $\alpha = 0.5$ | Fix $\alpha = 0.9$ | Fix $\alpha = 1.3$ |
|---|---|---|---|---|
| | $N(t)$ | $N(t)$ | $N(t)$ | $N(t)$ |
| 0 | 0 | 0 | 0 | 0 |
| 0.2 | 9.357284587 | -4177.748373 | 342.6586773 | -500.7257043 |
| 0.4 | 12.81476191 | -23104.48824 | 7969.477484 | -7271.787598 |
| 0.6 | 15.7046152 | -66819.99355 | 41813.93962 | -35593.96864 |
| 0.8 | 18.30995669 | -146638.5305 | 132250.6386 | -111081.8195 |
| 1 | 20.73658905 | -275685.3552 | 321797.0421 | -270748.1511 |
| 1.2 | 23.0376803 | -469308.4511 | 666514.8641 | -564463.4874 |
| 1.4 | 25.2443844 | -745501.5108 | 1237885.703 | -1056784.562 |
| 1.6 | 27.37679748 | -1125370.875 | 2125136.102 | -1829175.092 |
| 1.8 | 29.44877138 | -1633661.771 | 3438042.421 | -2982664.771 |
| 2 | 31.47033011 | -2299354.409 | 5310273.655 | -4641008.993 |
| 2.2 | 33.44900343 | -3156338.929 | 7903350.719 | -6954427.25 |
| 2.4 | 35.39061709 | -4244177.679 | 11411319.72 | -10104014.37 |
| 2.6 | 37.2997879 | -5608963.192 | 16066256.62 | -14306936.84 |
| 2.8 | 39.18024803 | -7304280.319 | 22144742.68 | -19822546.99 |
| 3 | 41.03506542 | -9392281.133 | 29975475.33 | -26959571.74 |
| 3.2 | 42.86679835 | -11944881.43 | 39948207.66 | -36084560.3 |
| 3.4 | 44.67760652 | -15045087.88 | 52524243.23 | -47631807.57 |
| 3.6 | 46.46933297 | -18788465.06 | 68248751.29 | -62115008.08 |
| 3.8 | 48.24356544 | -23284752.04 | 87765211.74 | -80140939.45 |
| 4 | 50.00168332 | -28659638.13 | 111832350.9 | -102425526.8 |

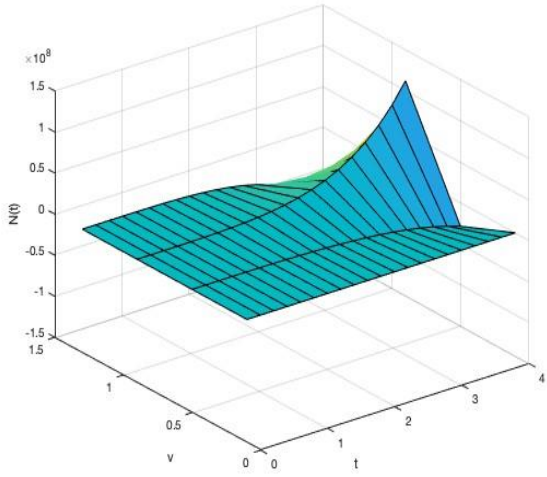 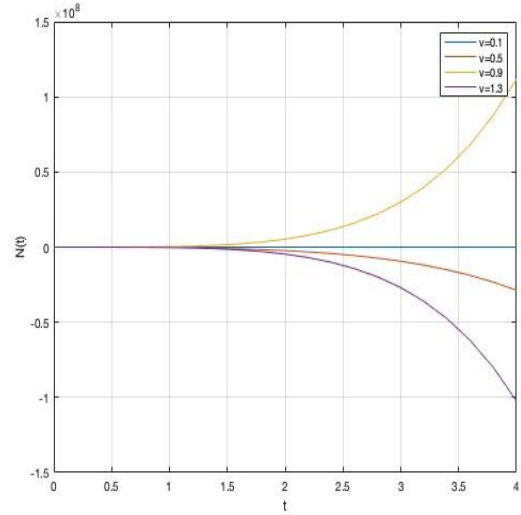

**Figure 2.2(a): 3D graph for (2.5)**             **2.2(b): 2D graph for (2.5)**

**Theorem-3:** If $\alpha > 0, c > 0, p\epsilon \mathbb{N}, k \in \mathbb{N}_0, A_{w,k} \geq 0, \mathbb{N}_0 = \mathbb{N} \cup \{0\}, 0 < \Re(\lambda) < 1$, then the solution of

$$N(t) - N_0 \left( {}_0D_t^\lambda \left( S_w^p(t) \right) \right) = -c^\alpha \, {}_0D_t^{-\alpha} \, N(t)$$

(2.6)

is given by

$$N(t) = N_0 \sum_{k=0}^{[w/p]} \frac{(-w)_{pk} \Gamma(\alpha k + 1)}{k!} A_{w,k} \, (t)^{k-\lambda} \, E_{\alpha, k-\lambda+1}(-c^\alpha \alpha)$$

(2.7)

**Proof:** Using Laplace Transform on both sides of (2.6) and in view of (1.5), we have

$$\mathcal{N}(s) = N_0 \sum_{k=0}^{[w/p]} \frac{(-w)_{pk} \Gamma(\alpha k + 1)}{k!} A_{w,k} \, (s)^{-(k-\lambda+1)} (1 + c^\alpha s^{-\alpha})^{-1}$$

$$= N_0 \sum_{k=0}^{[w/p]} \frac{(-w)_{pk} \Gamma(\alpha k + 1)}{k!} A_{w,k} \, (s)^{-(k-\lambda+1)} \sum_{n=0}^{\infty} (-c^\alpha s^{-\alpha})^n$$

$$= N_0 \sum_{k=0}^{[w/p]} \frac{(-w)_{pk} \Gamma(\alpha k + 1)}{k!} A_{w,k} \sum_{n=0}^{\infty} (-c^\alpha)^n (s)^{-(k-\lambda+1+\alpha n)}$$

Using inverse Laplace Transform, we have

$$N(t) = N_0 \sum_{k=0}^{[w/p]} \frac{(-w)_{pk} \Gamma(\alpha k + 1)}{k!} A_{w,k} \, (t)^{k-\lambda} \sum_{n=0}^{\infty} \frac{(-c^\alpha t^\alpha)^n}{\Gamma(k - \lambda + 1 + \alpha n)}$$

$$= N_0 \sum_{k=0}^{[w/p]} \frac{(-w)_{pk} \Gamma(\alpha k + 1)}{k!} A_{w,k} \, (t)^{k-\lambda} \, E_{\alpha, r-\lambda+1}(-c^\alpha t^\alpha)$$

**Numerical and Graphical Interpretation**

       By assuming $\alpha$ as constant and experimenting with different values of $t$, we get different values of $N(t)$, for (2.7), which are interpreted in the following table 2.3 and the graphs 3(a), 3(b) are presented to exhibit the behaviour of the result obtained.

**Table 2.3: The values of $N(t)$ with fix $\alpha$**

| $t$ | Fix $v = 0.1$ $N(t)$ | Fix $v = 0.5$ $N(t)$ | Fix $v = 0.9$ $N(t)$ | Fix $v = 1.3$ $N(t)$ |
|---|---|---|---|---|
| 0 | 0 | 0 | 0 | 0 |
| 0.2 | -1950.173536 | -4885.923448 | -11192.35049 | -16336.18258 |
| 0.4 | -6070.578434 | -44445.2887 | -116358.0826 | -162596.7718 |
| 0.6 | -18500.83829 | -233089.465 | -588435.9013 | -676299.129 |
| 0.8 | -53079.87818 | -868429.2226 | -2014166.963 | -1884214.866 |
| 1 | -141700.5816 | -2584153.229 | -5401711.834 | -4187368.689 |
| 1.2 | -352380.4475 | -6565812.952 | -12276453.94 | -8052581.738 |
| 1.4 | -820874.4438 | -14839290.44 | -24769392.61 | -14006701.48 |
| 1.6 | -1803194.599 | -30640481.55 | -45700168.29 | -22632305.83 |
| 1.8 | -3759145.864 | -58873939.25 | -78655920.79 | -34564329.49 |
| 2 | -7480367.024 | -106667609.1 | -128066769.4 | -50487315.8 |
| 2.2 | -14280446.61 | -184030298.3 | -199278470 | -71133117.57 |
| 2.4 | -26269536.11 | -304618118.3 | -298622660.1 | -97278933.61 |
| 2.6 | -46741575.73 | -486615800.9 | -433485006.3 | -129745604.7 |
| 2.8 | -80708851.13 | -753738502.5 | -612371502.7 | -169396115.3 |
| 3 | -135626183.2 | -1136359458 | -844973120.6 | -217134260.7 |
| 3.2 | -222355686.1 | -1672768630 | -1142228975 | -273903452.6 |
| 3.4 | -356432783 | -2410567296 | -1516388142 | -340685637.3 |
| 3.6 | -559705112.9 | -3408203363 | -1981070253 | -418500312.3 |
| 3.8 | -862428166.1 | -4736652003 | -2551324951 | -508403624.6 |
| 4 | -1305915023 | -6481246117 | -3243690308 | -611487541 |

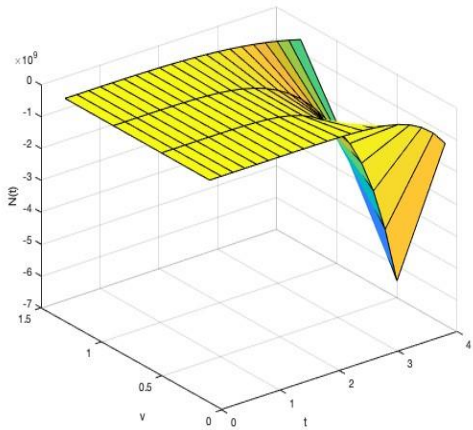
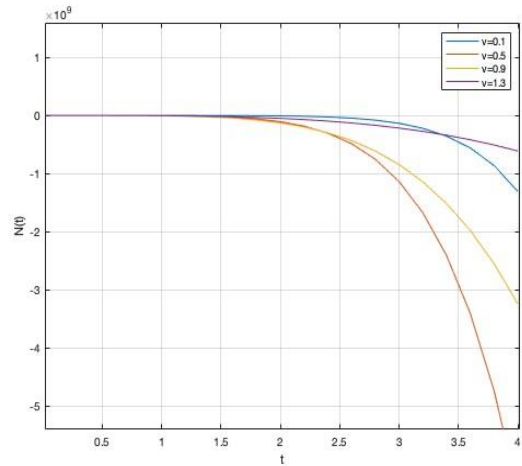

**Figure 2.3(a): 3D graph for (2.7)**            **2.3(b): 2D graph for (2.7)**

**Theorem-4:** If $\alpha > 0, c > 0, p \epsilon \mathbb{N}, k \in \mathbb{N}_0, A_{w,k} \geq 0, \mathbb{N}_0 = \mathbb{N} \cup \{0\}, 0 < \Re(\lambda) < 1$, then the solution of

$$N(t) - N_0 \left( {}_0D_t^\lambda \left( S_w^p(c^\alpha t^\alpha) \right) \right) = -d^\alpha {}_0D_t^{-\alpha} N(t)$$

(2.8)

is given as

$$N(t) = N_0 \sum_{k=0}^{[w/p]} \frac{(-w)_{pk} \Gamma(\alpha k + 1)}{k!} A_{w,k} (tc)^{\alpha k} t^{-\lambda} E_{\alpha, \alpha k - \lambda + 1}(-d^\alpha t^\alpha)$$

(2.9)

**Proof:** Using Laplace Transform on (2.8) and in view of (1.5), we have

$$\mathcal{N}(s) = N_0 \sum_{k=0}^{[w/p]} \frac{(-w)_{pk}\Gamma(\alpha k + 1)}{k!} A_{w,k}\, (c)^{\alpha k} s^{-(\alpha k - \lambda + 1)}(1 + d^\alpha s^{-\alpha})^{-1}$$

$$= N_0 \sum_{k=0}^{[w/p]} \frac{(-w)_{pk}\Gamma(\alpha k + 1)}{k!} A_{w,k}\, (c)^{\alpha k}\, s^{-(\alpha k - \lambda + 1)} \sum_{n=0}^{\infty} (-d^\alpha s^{-\alpha})^n$$

$$= N_0 \sum_{k=0}^{[w/p]} \frac{(-w)_{pk}\Gamma(\alpha k + 1)}{k!} A_{w,k}\, (c)^{\alpha k} \sum_{n=0}^{\infty} (-d^\alpha)^n (s)^{-(\alpha k - \lambda + 1 + \alpha n)}$$

Using inverse Laplace Transform, we have

$$N(t) = N_0 \sum_{k=0}^{[w/p]} \frac{(-w)_{pk}\Gamma(\alpha k + 1)}{k!} A_{w,k}\, (c)^{\alpha k} t^{\alpha k - \lambda} \sum_{n=0}^{\infty} \frac{(-d^\alpha t^\alpha)^n}{\Gamma(\alpha k - \lambda + 1 + \alpha n)}$$

$$= N_0 \sum_{k=0}^{[w/p]} \frac{(-w)_{pk}\Gamma(\alpha k + 1)}{k!} A_{w,k}\, (ct)^{\alpha k} t^{-\lambda}\, E_{\alpha,\alpha k - \lambda + 1}(-d^\alpha t^\alpha)$$

**Numerical and Graphical Interpretation**

By assuming $\alpha$ as constant and experimenting with different values of $t$, we get different values of $N(t)$, for (2.9), which are interpreted in the following table 2.4 and the graphs 4(a), 4(b are presented to exhibit the behaviour of the result obtained.

Table 2.4: The values of $N(t)$ with fix $\alpha$

| $t$ | Fix $v = 0.1$ $N(t)$ | Fix $v = 0.5$ $N(t)$ | Fix $v = 0.9$ $N(t)$ | Fix $v = 1.3$ $N(t)$ |
|---|---|---|---|---|
| 0 | 0 | 0 | 0 | 0 |
| 0.2 | 174.005681 | -43665.46423 | -3873.004587 | -6429.50019 |
| 0.4 | 87.00032928 | -204855.0131 | 29010.74341 | -86465.57532 |
| 0.6 | 57.99993493 | -586451.236 | 342766.5329 | -452309.0288 |
| 0.8 | 43.49987902 | -1311111.513 | 1452446.283 | -1493059.315 |
| 1 | 34.79987676 | -2523263.333 | 4191016.234 | -3790196.234 |
| 1.2 | 28.99988538 | -4389104.292 | 9747802.887 | -8129201.299 |
| 1.4 | 24.85703847 | -7096602.096 | 19701582.77 | -15508980.8 |
| 1.6 | 21.74990517 | -10855494.56 | 36049930.36 | -27149856.01 |
| 1.8 | 19.33324684 | -15897289.6 | 61235774.02 | -44500536.05 |
| 2 | 17.39992046 | -22475265.25 | 98171771.8 | -69244326.54 |
| 2.2 | 15.81810757 | -30864469.64 | 150262926.9 | -103304740.6 |
| 2.4 | 14.49992882 | -41361721.03 | 221427745.8 | -148850628.4 |
| 2.6 | 13.38454404 | -54285607.76 | 316118164.9 | -208300909.1 |
| 2.8 | 12.42849524 | -69976488.31 | 439338420.1 | -284328969 |
| 3 | 11.59991234 | -88796491.23 | 596662996.9 | -379866774.9 |
| 3.2 | 10.8748918 | -111129515.2 | 794253771.2 | -498108739.8 |
| 3.4 | 10.23515339 | -137381229 | 1038876432 | -642515374.6 |
| 3.6 | 9.666478053 | -167979071.6 | 1337916258 | -816816747.9 |
| 3.8 | 9.157639097 | -203372251.9 | 1699393319 | -1025015778 |
| 4 | 8.699654031 | -244031749.1 | 2131977146 | -1271391372 |

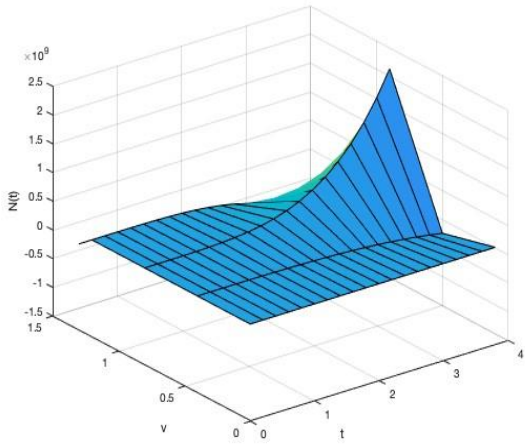 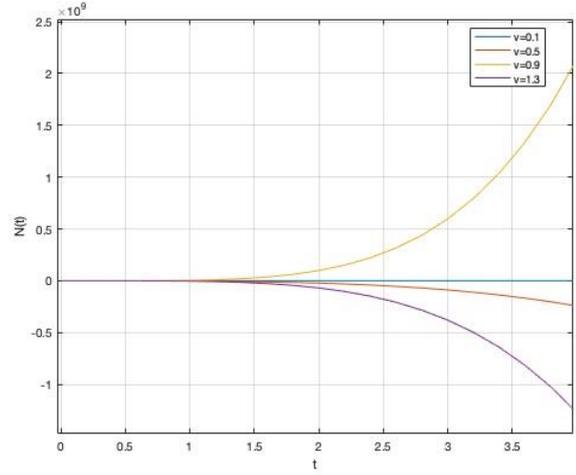

**Figure 2.4(a): 3D graph for (2.9)**  2.4(b): 2D graph for (2.9)

## 3. Particular Cases

(i) Taking $d = c$ in (2.4) then equation reduces as
$$N(t) - N_0\{ S_w^p(c^\alpha t^\alpha)\} = -c^\alpha {}_0D_t^{-\alpha} N(t)$$
(3.1)

and the outcome is given by
$$N(t) = N_0 \sum_{k=0}^{[w/p]} \frac{(-w)_{pk}\Gamma(\alpha k + 1)}{k!} A_{w,k} (ct)^{\alpha k} E_{\alpha,(\alpha k+1)}(-c^\alpha t^\alpha)$$
(3.2)

(ii) Taking $c = 1$ in (2.4) then equation reduces as
$$N(t) - N_0\{ S_w^p(t^\alpha)\} = -d^\alpha {}_0D_t^{-\alpha} N(t)$$
(3.3)

and the outcome is given by
$$N(t) = N_0 \sum_{k=0}^{[w/p]} \frac{(-w)_{pk}\Gamma(\alpha k + 1)}{k!} A_{w,k} (t)^{\alpha k} E_{\alpha,(\alpha k+1)}(-t^\alpha)$$
(3.4)

(iii) Taking $d = c$ in (2.8) then equation reduces as
$$N(t) - N_0 \left( {}_0D_t^\lambda \left( S_w^p(c^\alpha t^\alpha) \right) \right) = -c^\alpha {}_0D_t^{-\alpha} N(t)$$
(3.5)

and the outcome is given by
$$N(t) = N_0 \sum_{k=0}^{[w/p]} \frac{(-w)_{pk}\Gamma(\alpha k + 1)}{k!} A_{w,k} (tc)^{\alpha k} t^{-\lambda} E_{\alpha,\alpha(\mu+k)-\lambda+1}(-c^\alpha t^\alpha)$$
(3.6)

(iv) Taking $c = 1$ in (2.8) then equation reduces as
$$N(t) - N_0 \left( {}_0D_t^\lambda \left( S_w^p(t^\alpha) \right) \right) = -d^\alpha {}_0D_t^{-\alpha} N(t)$$
(3.7)

and the outcome is given by
$$N(t) = N_0 \sum_{k=0}^{[w/p]} \frac{(-w)_{pk}\Gamma(\alpha k + 1)}{k!} A_{w,k} (t)^{\alpha k} t^{-\lambda} E_{\alpha,\alpha(\mu+k)-\lambda+1}(-t^\alpha)$$
(3.8)

## Conclusion

Due to the usefullness and a great importance FKE in a variety of applied sciences and engineering fields as well, numerous studies have been conducted in this area has been executed. We have developed the four theorems in this study include a solution using Laplace Transform to a generalised form of FKE involving Srivastava polynomial. Our primary findings, which are broad in nature, produce a wealth of novel and well-known conclusions in terms of basic functions when appropriate parametric restrictions are met. These findings could find extensive application in a variety of scientific and technological domains. Furthermore, given the close relationships of the Srivastava polynomial, it appears to be rather easy to create other known and unutilized fractional kinetic equations. Therefore, the analysis presented in this paper would immediately yield a number of results about various unique functions occurring in the fields of engineering, astronomy, and scientific mathematics.

**Conflicts of interest:** None